# Low Voltage Ride Through (LVRT) Constrained Transient Stability Assessment Using Lyapunov Functions Family Method

Chen Wang
Chetan Mishra
Dominion Energy
Richmond, Virginia 23220, USA

Virgilio A. Centeno
Bradley Department of Electrical and Computer Engineering,
Virginia Polytechnic Institute and State University
Blacksburg, Virginia 24061, USA

*Abstract*—Growing penetration of renewable generation in power systems brings more challenges in transient stability assessment using direct methods. One of the reasons is the inability to assess the risk of instability brought by tripping of a large amount of RGs due to violation of their ride through curves. In this paper, a scalable approach is proposed based on the Lyapunov functions family to estimate the stability region under ride through constraints. An inner polytopic approximation to the feasibility region is proposed to convexify the overall problem. The acquired constrained stability region estimate helps capture trajectories that trigger undesirable tripping of renewable generation. A 2-machine system is used to visualize its effectiveness.

## I. Introduction

Transient stability assessment (TSA) refers to estimating the maximum fault clearing time such that the system remains stable (critical clearing time, or CCT). Conventional direct methods [1] do so by estimating the whole or relevant portion of the stability region (SR) [2] of the post-fault system and then simply checking whether the system at the time of fault clearing is within that region. The purpose is to capture instability phenomena like loss of synchronism, voltage collapse, etc which is characterized by the fault on trajectory exiting the SR of the post fault stable equilibrium point (SEP).

Renewable generation (RG), specifically at distribution level, is made to be tripped offline during system disturbances to prevent feeding in an islanding situation. This is done with the help of ride-through settings which are limits on operating conditions mainly voltage and frequency. LVRT, which gives the lower limit on voltage, is the most relevant one when studying faults. A major issue is that the trip logic is based on local measurements of voltage and frequency at the point of interconnection which may not correctly observe the islanding event. There can be disturbances that may not result in islanding but cause wide-spread tripping of RGs thereby causing a system collapse. This is characterized by the system trajectory exiting the feasibility region (FR). This region refers to the points in state space where all operating constraints are met. The traditional direct methods cannot capture this phenomenon.

The SR characterization for constrained dynamical systems was introduced in [3]. Mishra et.al. proposed using sum of squares programming [4], [5] to estimate this region for direct stability assessment of power systems under LVRT constraints. However, this approach faces serious computational limitations and therefore not scalable to large systems. Alberto et.al [6] proposed using the Potential Energy Boundary Surface (PEBS) method for TSA of such LVRT constrained systems which is known to have reliability concerns when used for multi-machine systems [7]. Turitsyn et.al [8] proposed an approach that uses a family of Lyapunov functions (LFs) sharing the same structure as the energy function to get a decent estimate of the SR. This approach is scalable to large systems. We will be exploiting this approach along with advances in convex optimization for estimating the SR considering constraints based on the LVRT curves. This region is referred to as the LVRT constrained stability region (CSR). Some of the other important contributions of this work include,

1. Formulation of Lyapunov Functions Family (LFF) approach in the center of angle (COA) reference frame.
2. Optimal polytopic inner approximation of LVRT constrained FR for convexifying the feasibility validation process within the estimation.

The paper is organized as follows: section II talks about the CSR concept and LFF method to estimate the LVRT CSR; section III talks about polytope approximation of the LVRT constrained FR; the effectiveness of the proposed CSR estimation method is demonstrated on a 2-machine 3-bus test case in section IV; section V gives conclusions.

## II. Constrained Transient Stability Assessment with Lyapunov Functions Family

### A. Constrained Stability Region

By definition, CSR of a given SEP, $x_{sep}$ of a system is a set of points from which the emerging trajectories never exit the FR and converge to $x_{sep}$. The difference between CSR and the overlapping area of SR with FR can be remarkable. For example in Figure 1, even though $x_0^{us}$ is inside the FR and SR, the trajectory starting from it crosses the feasibility boundary (FB) eventually. Thus, this point is not in the CSR (shaded region). Furthermore, the FB is comprised of three types of components: the flow-out part (dashed line) where trajectories starting in the vicinity cross the boundary to the infeasible region, the flow-in part (straight line) where they get repelled to the feasible side and the semi-saddle points where they are tangential to the FB. Ref [3] shows CSR can only take flow-in and semi-saddle components of the FB as parts of its boundary.

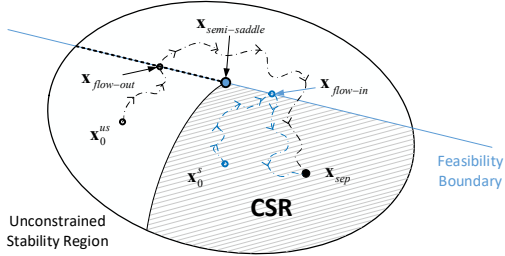

Figure 1. Constrained Stability Region Example for 2-dimensional System

Now, the complexity of the CSR that can be estimated using only a level set of a given LF $V(x)$ is heavily restricted by the degree of the function chosen. This usually introduces conservativeness which can be reduced by using a combination of parts of the FB itself and a level set of the LF.

The aim is to find the maximal level set $\Theta = \{x | V(x) < v_{max}\}$ satisfying the following constraints –

$$\begin{cases} \Theta \cap \partial\Gamma \subseteq \partial\Gamma_{flow-in} \cup \partial\Gamma_{semi-saddle} \\ \Theta \subseteq \{x | \dot{V}(x) < 0\} \\ x_{sep} = \Theta \cap \{x | \dot{V}(x) = 0\} \end{cases} \quad (1)$$

where, $\Gamma$ refers to the FR and $\partial\Gamma$ is its boundary, FB; $\partial\Gamma_{flow-in}$ and $\partial\Gamma_{semi-saddl}$ refers to the flow-in and semi-saddle components on the FB, respectively. The CSR estimate is then given by $\Theta \cap \Gamma$. How to estimate this region using LFF is explained in the next section.

### B. Finding Candidate Lyapunov Functions

As mentioned before, The LFF method essentially searches for optimal parameter values for a function belonging to the same family of functions to which the analytical energy function [1] belongs. In [8], the method was described in the absolute angle frame. A major issue with analyzing the power system model in absolute angles is that the vector field itself is a function of differences of angles which means if $(\delta^*, 0)$ is an equilibrium point, so is $(\delta^* + constant, 0)$ yielding a single dimensional boundary-less manifold of equilibrium points having the exact same characteristics. From the power systems point of view, all of these points result in the same operating conditions which is effectively defined in terms of power flows. This means that the CSR needs to be estimated for a boundary-less surface of desired SEPs instead of a single one. Furthermore, using an LF in which, as done in [8], the angle terms are defined not just in angle differences will have a level set having a closure and therefore unable to contain that whole surface of desirable SEPs. Thus, we propose a formulation of the LFF approach in the widely known center of angle reference (COA) frame in which the surface of SEPs collapses to a single isolated point.

The model for a lossless power system network reduced model with classical generators is shown below in COA frame with uniform damping. RGs are modeled as negative loads.

$$\dot{\omega}_k^{coa} = \frac{P_{m_k}}{m_k} - \frac{\sum_i P_{m_i}}{\sum_i m_i} - \frac{\sum_j B_{kj} E_k E_j \sin(\delta_k^{coa} - \delta_j^{coa})}{m_k} - \left( \frac{-\sum_i \sum_j B_{ij} E_i E_j \sin(\delta_i^{coa} - \delta_j^{coa})}{\sum_i m_i} \right) - \frac{d_k}{m_k} \omega_k^{coa} \quad (2)$$

where $m_k$ is the inertia constant of the $k$-th synchronous generator in the system; $P_{m,k}$ is its mechanical input power; $\delta_k^{coa}$ is its corresponding rotor angle and $\omega_k^{coa}$ angular velocity, both in the COA reference frame, which satisfy $\sum_i m_i \delta_i^{coa} = 0$ and $\sum_i m_i \omega_i^{coa} = 0$ ; $d_k$ is its damping; $B_{kj}$ is the corresponding entry of the Kron-reduced post-fault system susceptance matrix; $E_k$ is its internal voltage magnitude.

As in [8], $x$ is defined as the deviation from the SEP. Here however it is in COA reference frame

$$\begin{cases} x_1 = [(\delta_1^{coa} - \delta_1^{coa*}) \quad \cdots \quad (\delta_n^{coa} - \delta_n^{coa*})]^T \\ x_2 = [(\omega_1^{coa} - \omega_1^{coa*}) \quad \cdots \quad (\omega_n^{coa} - \omega_n^{coa*})]^T \end{cases} \quad (3)$$

where, $\delta_k^{coa*}$ denotes the rotor angle value of the $k$-th generator in COA frame at the SEP.

Assuming the mechanical input power is constant, and $\delta_{ij} = \delta_i - \delta_j$, the state equation (2) can be written as

$$\dot{\omega}_k^{coa} - \dot{\omega}_k^{coa*} = \left( \frac{\sum_i \sum_j B_{ij} E_i E_j (\sin \delta_{ij}^{coa} - \sin \delta_{ij}^{coa*})}{\sum_i m_i} \right) - \frac{\sum_j B_{kj} E_k E_j (\sin \delta_{kj}^{coa} - \sin \delta_{kj}^{coa*})}{m_k} - \frac{d_k}{m_k} \omega_k^{coa} \quad (4)$$

For convenience, in the rest of the paper angles in COA frame will be denoted by $\delta_i$. Take the states to be $x = [x_1 \quad x_2]^T$. Separating the linear and non-linear parts, the state space model is of the form

$$\dot{x} = Ax - BF(Cx) \quad (5)$$

where $F(Cx) = \left[ (\sin \delta_{kj} - \sin \delta_{kj}^*)_{\{k,j\} \in \varepsilon} \right]^T$ which is the non-linear part; $Cx$ indicates the angle differences among generators; $\varepsilon$ represents the set of branches in the Kron-reduced model. The parameter matrices of the model are as follows:

$$A = \begin{bmatrix} 0_{n \times n} & I_{n \times n} \\ 0_{n \times n} & -M^{-1}D \end{bmatrix}, B = -\gamma^{(1)} + \gamma^{(2)} \quad (6)$$

$$\gamma^{(1)} = \begin{bmatrix} \frac{1}{\sum_i m_i} B_{12} E_1 E_2 & \cdots & \frac{1}{\sum_i m_i} B_{kj} E_k E_j & \cdots \\ \vdots & & \vdots & \\ \frac{1}{\sum_i m_i} B_{12} E_1 E_2 & \cdots & \frac{1}{\sum_i m_i} B_{kj} E_k E_j & \cdots \\ \vdots & & \vdots & \end{bmatrix}_{n \times |\varepsilon|} \quad (7)$$

$$\gamma^{(2)} = i^{th} node \begin{bmatrix} \vdots \\ 0 \cdots 0 & \frac{1}{m_i} B_{i1} E_i E_1 & \cdots & \frac{1}{m_i} B_{in} E_i E_n & 0 \cdots 0 \\ \vdots \end{bmatrix}_{n \times |\varepsilon|} \quad (8)$$

where, $n$ is the number of synchronous generators; $|\varepsilon|$ refers to the number of branches in the system. For a lossless system, the summation of the electric power in the system should be zero. Thus the corresponding term $\gamma^{(1)} F(Cx) = 0$.

Energy function [1] for the above system belongs to the following family of functions,

$$V(x) = \frac{1}{2} x^T Q x - \sum_{\{k,j\} \in \varepsilon} K^{\{k,j\}} \left( \cos \delta_{kj} + \delta_{kj} \sin \delta_{kj}^* \right) \quad (9)$$

where, $Q$ and $K$ are the two parameter matrices; $Q$ is symmetric positive definite; $K$ is diagonal positive definite and $K^{\{k,j\}}$ is its diagonal entry corresponding with the branch between bus $k$ and bus $j$. By La Salle Invariance Principle [9] we search for these parameter values such that in some region containing the SEP,

$$\dot{V}(x) = x^T QAx - x^T QBF + (Gx)^T KF \leq 0 \quad (10)$$

where, $Gx = \begin{bmatrix} \cdots & \dot{\delta}_{kj} & \cdots \end{bmatrix}$ and $F = F(Cx)$.

It was shown in [8] that when $|\delta_{kj} + \delta_{kj}^*| \leq \pi$,

$$(\sin \delta_{kj} - \sin \delta_{kj}^*)^2 \leq (\delta_{kj} - \delta_{kj}^*)(\sin \delta_{kj} - \sin \delta_{kj}^*) \quad (11)$$

Using the S-procedure[10] and this inequality, the problem of finding $Q$ and $K$ can be written as a linear matrix inequality (LMI) as shown in [8]:

$$\begin{bmatrix} A^TQ + QA & -(QB - (KCA)^T - C^TH) \\ -(QB - (KCA)^T - C^TH)^T & -2H \end{bmatrix} \leq 0 \quad (12)$$

where, $H$ is also a positive diagonal matrix. Once a candidate set of parameter matrices is estimated giving an LF, the CSR will be estimated using a level set of this function in combination with parts of the FB.

### C. Estimating Constrained Stability Region

This candidate level set $\{(x_1, x_2)|V(x_1, x_2) \leq v\}$ has to satisfy the conditions given in Section II.A in order to be considered as an estimate for the CSR. We know that $\dot{V} \leq 0$ inside $\{|\delta_{kj} + \delta_{kj}^*| \leq \pi \, \forall k, j\}$ which is a polytope. Thus, as long as the trajectories are contained inside this set, $V$ will continue decreasing till it reaches the largest invariant set contained in the set $\{x|\dot{V}(x) = 0\}$ by La Salle Invariance Principle. As long as this set only contains a single SEP, it is guaranteed that the trajectories will converge to it.

In the present work, the only constraint we dealt with is the LVRT constraint which forms the FB. By definition, it is a lower limit on the bus voltage magnitude at the point of interconnection of RGs. For a classical system model, it can be represented as a nonlinear function of the rotor angles, which will be shown in the next section. For the time being, let us assume that we can find a polytopic inner approximation to the LVRT constrained FR. Thus, the intersection of the approximate LVRT constrained FR with the region where $\dot{V} \leq 0$ given by $\{|\delta_{kj} + \delta_{kj}^*| \leq \pi \, \forall k, j\}$ can be written in the form $\{x|L_{ineq}x_1 + l_{ineq} \leq 0\}$. Here, $L_{ineq}$ is a coefficient matrix and $l_{ineq}$ is a constant vector both with $N_{constraint}$ number of rows. We will refer to this region as the approximate combined feasibility region (ACFR) which is also an invariant set. The boundary of this set referred to as ACFB is a collection of hyperplanes making up facets of the polytope. In general, the $i^{th}$ facet of this polytope can be uniquely described by the set of equations $\{x|L_{ineq}^{(i)}x_1 + l_{ineq}^{(i)} = 0, L_{ineq}^{(j)}x_1 + l_{ineq}^{(j)} \leq 0 \, \forall j \neq i\}$ where $L_{ineq}^{(i)}$ refers to the $i^{th}$ row of $L_{ineq}$.

The time derivative at a point lying on the $i^{th}$ facet of the ACFB can simply be written as $L_{ineq}^{(i)}x_2$. The sign of this derivative can then be used to classify that point as flow-in, flow-out or semi-saddle.

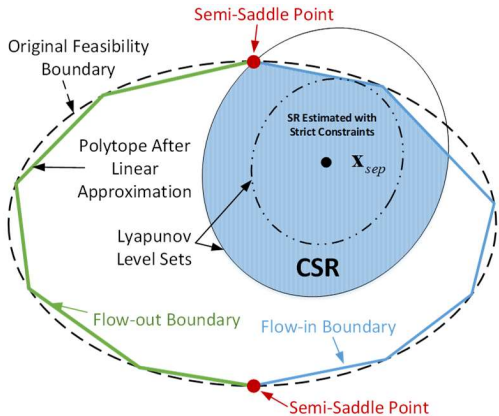

Figure 2. CSR Estimated With Polytopic Feasibility Region

In order to check if the intersection of a Lyapunov level set $\{(x_1, x_2)|V(x_1, x_2) \leq v\}$ with the ACFR can be used as an estimate for the CSR, we simply need to make sure that no flow-out points on the ACFB lie inside it. Checking if any flow-out points on the $i^{th}$ facet lie inside this set can be formulated as an optimization problem,

$$\begin{aligned} &\exists \ x_1, x_2 \\ s.t. \quad &max_{x_1,x_2} L_{ineq}^{(i)} x_2 > 0 \\ &L_{ineq}^{(i)} x_1 + l_{ineq}^{(i)} = 0 \\ &L_{ineq}^{(j)} x_1 + l_{ineq}^{(j)} \leq 0, \forall j \neq i \\ &L_{eq}x + l_{eq} = 0 \\ &V(x_1, x_2) \leq v \end{aligned} \quad (13)$$

If a solution exists to this problem, there exists at least one flow-out point making the candidate estimate of CSR $\{(x_1, x_2)|V(x_1, x_2) \leq v, L_{ineq}x_1 + l_{ineq} \leq 0, L_{eq}x + l_{eq} = 0\}$ non-invariant and thus invalid. This check can be done in a parallel manner across all the facets of the ACFB. One thing to mention here is that the equality constraint $L_{eq}x + l_{eq} = 0$ comes from the fact that in the COA frame, the system dynamics are evolving on a lower-dimensional manifold as mentioned before which can be explicitly written as $\sum_i m_i \delta_i^{coa} = M \times x_1 = 0$ and $\sum_i m_i \omega_i^{coa} = M \times x_2 = 0$.

Now, the above problem is a non-convex non-linear optimization problem due to non-convexity of the constraint $V(x_1, x_2) \leq v$. However, if the operating domain is tightened to $|\delta_{kj}^{coa}| \leq \frac{\pi}{2}$, $V(x_1, x_2)$ is convex[8]. Therefore, this replaces the original constraint $|\delta_{kj} + \delta_{kj}^*| \leq \pi$ with the definition of ACFR now revised accordingly. For each candidate LF ($Q_i, K_i$ pair), a linear search of the level set value $v$ is expanded starting from 0 until a solution exists for (13). Figure 2 gives a visual representation of the overall process.

A fault will be called stable if the state at the time of fault-clearing, $x_0$ lies inside the estimated CSR. However, being outside does not necessarily mean it is unstable, as it could be due to conservativeness in CSR estimation. An iterative approach to evolve the Lyapunov function to reduce the likelihood of such scenarios was proposed in [8] which we use here. The improvement is assured in the estimate of CSR in terms of the amount of overlap with the portion of the state space relevant to a given fault trajectory under study.

### III. POLYTOPIC INNER APPROXIMATION OF LVRT CONSTRAINED FEASIBILITY REGION

As discussed in the previous section, a polytopic inner approximation of ACFR which is the intersection of LVRT constrained FR with $|\delta_{kj}^{coa}| \leq \frac{\pi}{2}$ needs to be found. The general form of LVRT constraints is

$$|v_k^{(RG)}| \geq V_{LVRT}(t) \quad (14)$$

where, $|v_k^{(RG)}|$ represents the voltage magnitude of bus $k$, which is the interconnection bus of an RG; $V_{LVRT}(t)$ represents the time-varying LVRT curve value. Time-varying constraints are difficult to handle in Lyapunov approach. Therefore we conservatively convert it to the following time-independent formulation.

$$|v_k^{(RG)}| \geq LVRT_{max} = \max_t V_{LVRT}(t) \quad (15)$$

This approach only brings conservativeness in systems with slow voltage recovery and usually not so much in systems where loads are modeled as static. Using the Kron-reduced model, one can find the relation between the aforementioned states and the bus voltage as follows [4]

$$\begin{bmatrix} I_g \\ 0 \end{bmatrix} = Y_{ext} \begin{bmatrix} E \\ v \end{bmatrix} \quad (16)$$

where, $I_g$ refers to the generator terminal current vector; $Y_{ext}$ represents the extended admittance matrix [11], where all the loads and generators' transient reactances are transferred into branch admittances; $E$ refers to the generator internal EMF vector; and $v$ is the vector of bus voltages. So an equation between $E$ and $v$ can be found using the second row in (16):

$$v = P \cdot E \quad (17)$$

where, $P$ is the parameter matrix derived from $Y_{ext}$. Furthermore, one can find the square of bus voltage magnitude as a function of generator rotor angles making the LVRT constraint as:

$$g_k(\delta) = |v_k|^2 = \sum_i \sum_j C_i C_j \cos(\delta_{ij} + \delta_{ic} - \delta_{jc}) \geq LVRT_{max}^2 \quad (18)$$

where, $g_k$ is the function of the rotor angles vector, $\delta$, corresponding to RG integrated bus $k$; $C_i$ and $\delta_{ic}$ represent the magnitude and phase angle constants of bus $k$, which are defined as follow:

$$C_i = E_i |p_{k,i}|, \delta_{ic} = ang(p_{k,i}) \quad (19)$$

and here $E_i$ refers to the magnitude of the internal voltage of generator $i$; $p_{k,i}$ refers to the entry of $P$ in the position of row $k$ and column $i$. This region is non-convex by nature.

Since the LVRT constraint is a summation of cosine terms in angle differences with each angle difference pair constrained to the finite region $|\delta_{kj}| \leq \frac{\pi}{2}$, we can simply find a lower linear approximation individually for each term and add them together to convert the overall non-linear constraint to a linear one. That being said, this is bound to bring conservativeness in the inner approximation of the LVRT constrained FR. This can be overcome by approximating each term with a combination of linear functions as shown in Figure 3 (a), (b), and (c).

As an example, assume the LVRT constraint is $g(y) = c_1 \cos(y_1) + c_2 \cos(y_2) \geq \lambda$. The first term is approximated by the minima of two linear functions, i.e. $\min(a_{11}y_1 + b_{11}, a_{12}y_1 + b_{12})$. Similarly, The second term is approximated by $\min(a_{21}y_2 + b_{21}, a_{22}y_2 + b_{22})$. The overall FR approximation is written as $g^{approx}(y) = \min(a_{11}y_1 + b_{11}, a_{12}y_1 + b_{12}) + \min(a_{21}y_2 + b_{21}, a_{22}y_2 + b_{22}) \geq \lambda$.

This region can also be written by the following set of constraints yielding a polytope – $\{a_{11}y_1 + b_{11} + a_{21}y_2 + b_{21} \geq \lambda, a_{12}y_1 + b_{12} + a_{21}y_2 + b_{21} \geq \lambda, a_{11}y_1 + b_{11} + a_{22}y_2 + b_{22} \geq \lambda, a_{12}y_1 + b_{12} + a_{22}y_2 + b_{22} \geq \lambda\}$. Generalizing, an LVRT constraint with $N_{cos}$ cosine terms where the $i^{th}$ term is approximated by the minimum of $N_i$ linear functions will result in an overall polytope with $\prod_{i=1}^{N_{cos}} N_i$ facets/linear constraints.

Let us assume that for a given cosine term $\cos(\delta_{ij} + \delta_{ijc})$ in (18), we need to find this set of lower bounding $N_{line}$ number of linear functions of the form $a_k \times \delta_{ij} + b_k$ for $k = 1: N_{line}$. The approximate function as mentioned before is formulated as $\min_k(a_k \times \delta_{ij} + b_k)$. Earlier we saw that $|\delta_{kj}| \leq \frac{\pi}{2}, \forall k, j$ for convexity of LF $V$ which restricts our analysis to this region. The problem of finding these functions can be formulated as the following optimization problem which can be solved in parallel across all the $(i, j)$ pairs.

$$\min_{[a_1, a_2 \ldots],[b_1, b_2 \ldots]} \sum_p \left( \cos(\delta_{ij}(p) + \delta_{ijc}) - \min(\min_k(a_k \times \delta_{ij}(p) + b_k)) \right)^2$$
$$s.t. \quad \cos(\delta_{ij}(p) + \delta_{ijc}) \geq \min\left(\min_k(a_k \times \delta_{ij}(p) + b_k)\right)$$
$$|\delta_{ij}(p)| \leq \frac{\pi}{2}$$
(20)

Directly using $a, b$ as decision variables results in a space with a large number of local minima. Solutions where the lower bounding approximating function should have intersected the original function but did not were common and evidently suboptimal as shown in Figure 3 (d). Thus, we restricted our search space by only looking for resulting polytopes having vertices on the actual curve. Since the function is in a single variable ($\delta_{ij}$), we chose $N_{line} + 1$ number of points, $\{(\delta_{ij}, \cos(\delta_{ij} + \delta_{ijc}))\}$, on the original curve and sort them by their $\delta_{ij}$ coordinate. The lines connecting successive points become the approximating linear functions/faces of the final polytope. Thus, the search is now for an optimal set of points on the original curve. The conservativeness of the ACFR, when compared to the actual combined FR, decreases with increasing $N_{line}$ as shown in Figure 3 (e) and (f). For each RG integrated bus, its approximated LVRT constraints have the linear form as

$$g^{approx.} = aLx_1 + \sum b + C_{sep} > LVRT_{max}^2 \quad (21)$$

where $L$ maps rotor angles to angle differences, and we have $Lx_1 = \left[ \left( \delta_i - \delta_j - (\delta_i^* - \delta_j^*) \right)_{\{i,j\} \in \varepsilon} \right]^T; C_{sep} = aL\delta^*$.

## IV. CASE STUDY

The study case used to demonstrate the effectiveness of the proposed method is a 2-machine 3-bus system. The network model can be seen in Figure 4. The reason for using a 2-machine system is that in the COA frame the state space becomes 2D making it possible to visualize the CSR. The method can be easily implemented on larger-scale systems because of the efficiency advantage brought by the LMI used for LF based direct methods for TSA and the polytopic approximation of the FR. $LVRT_{max}$ is set to be 0.85. The fault being assessed is a three-phase to ground fault between bus 1 and bus 2. The post-fault initial state $x_0$ used for evolving LF as mentioned in the last paragraph of section II.C is the state value reached on sustaining the fault for 0.2 seconds.

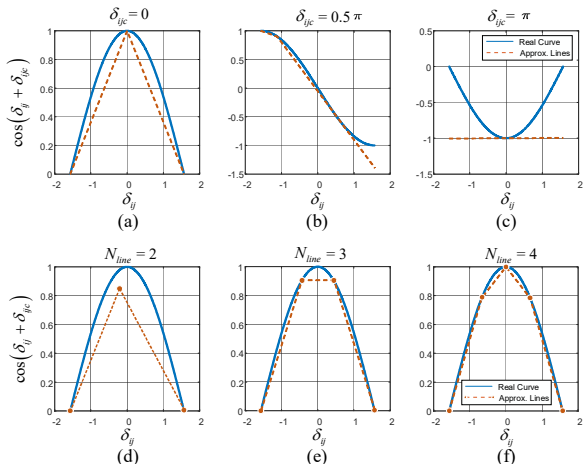

Figure 3. Examples of Linear Approximation of Single Cosine Curve

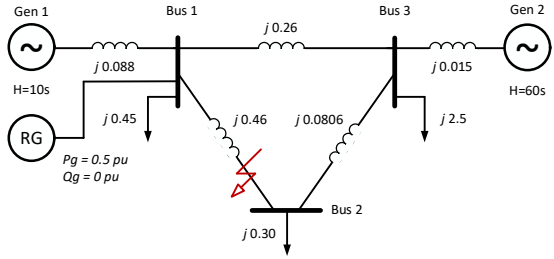

Figure 4. 2-machine 3-bus Study Case Diagram

The CSR estimation results can be seen in Figure 5. The orange arrows show the vector field of the post fault system at each point. Also shown are the boundaries of the original (brown lines) and the polytopic approximation (black lines) of the LVRT constrained FR, as well as the region for convexity of $V$ (light green lines). Here it can be seen that the ACFR estimate covers a large portion of the original combined FR. The flow out portions of the corresponding ACFB are marked in grey dashed lines, which can be seen from the arrow directions of the vector field.

We start by using the energy function as the initial LF $V_{initial}(x)$. As we start expanding its level set, the expansion stops as soon as it hits the edge of the flow out portion on the right, yielding the region marked in blue. This is popularly known as the closest UEP method [1]. Since $x_0$ does not fall inside it, we continue to check whether this could be due to the conservativeness in the estimated CSR. For that, we again search for a new LF candidate $V^{(2)}(x)$. The maximum level-set (red circle) of $V^{(2)}(x)$ is seen to contain $x_0$. This means that the system is stable for the particular disturbance. $V^{(2)}(x)$ is thus less conservative than the closest UEP method.

Let us now compare the estimated CSR (red dot-shaded region) with the actual CSR (blue dot-shaded region) in Figure 6. It can be seen that the estimated CSR covers a significant portion of the actual CSR, thereby demonstrating sthe effectiveness of our approach.

## V. CONCLUSIONS

In this paper, the stability region of a power system under LVRT constraints is estimated using Lyapunov's direct method. First, a family of candidate Lyapunov functions is defined in the COA frame, within which we search for a good Lyapunov function candidate. The problem of estimating the stability region under LVRT constraints using the found

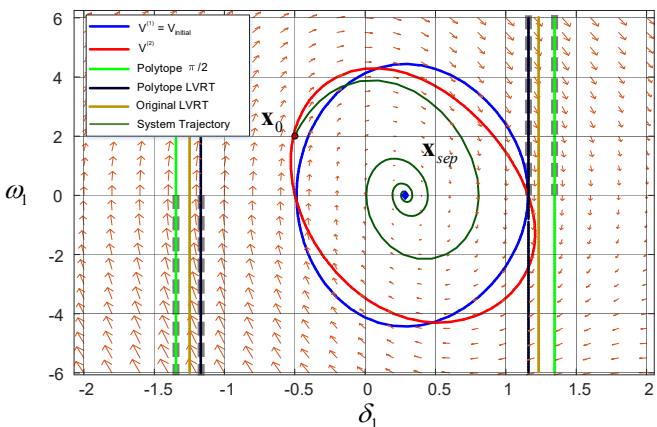

Figure 5. Estimated Constrained Stability Region

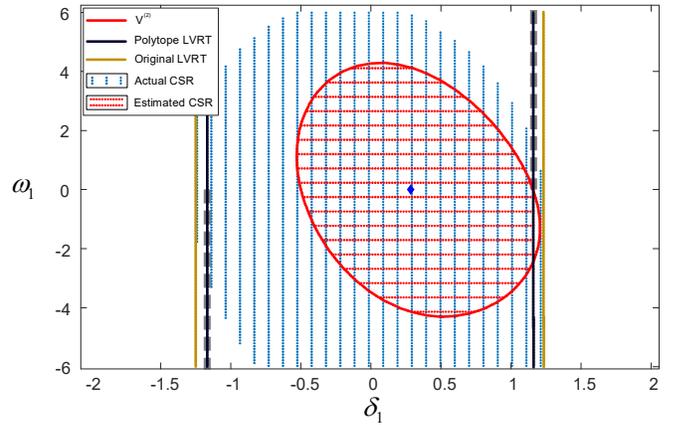

Figure 6. Comparison between Actual and Estimated CSR

Lyapunov function is non-convex by nature. Therefore, in order to convexify it for large-scale system applications, a methodology for obtaining an optimal polytopic inner approximation of the feasibility region is proposed. The method's effectiveness is demonstrated on a 2-machine 3-bus system. The proposed method is a good first step for quickly capturing the disturbances that can initiate wide-spread tripping of renewable generation, thereby increasing the threat of instability in power systems with high renewable penetration. The proposed method is shown to have less conservativeness than the closest UEP method. It was also seen to provide a reasonably good estimate of the stability region. The efficiency of the method due to linearity also allows its application to large-scale systems which will be explored in future work.

REFERENCES

[1] H. Chiang, *Direct Methods for Stability Analysis of Electric Power Systems: Theoretical Foundation, BCU Methodologies, and Applications*. John Wiley & Sons, Ltd, 2010.
[2] C. Mishra, J. S. Thorp, V. A. Centeno, and A. Pal, "Estimating Relevant Portion of Stability Region using Lyapunov Approach and Sum of Squares," in *2018 IEEE Power Energy Society General Meeting (PESGM)*, 2018, pp. 1–5.
[3] K. L. Praprost and K. A. Loparo, "A stability theory for constrained dynamic systems with applications to electric power systems," *IEEE Trans. Autom. Control*, vol. 41, no. 11, pp. 1605–1617, Nov. 1996.
[4] C. Mishra, J. S. Thorp, V. A. Centeno, and A. Pal, "Stability region estimation under low voltage ride through constraints using sum of squares," in *2017 North American Power Symposium (NAPS)*, 2017, pp. 1–6.
[5] C. Mishra, A. Pal, J. S. Thorp, and V. A. Centeno, "Transient Stability Assessment of Prone-to-Trip Renewable Generation Rich Power Systems Using Lyapunov's Direct Method," *IEEE Trans. Sustain. Energy*, vol. 10, no. 3, pp. 1523–1533, Jul. 2019.
[6] A. P. Sohn and L. F. C. Alberto, "Stability analysis of a wind power system via PEBS method," in *2017 IEEE Power Energy Society General Meeting*, 2017, pp. 1–5.
[7] H. Chiang, F. F. Wu, P. P. Varaiya, and C. Tan, "Theory of the potential energy boundary surface," in *1985 24th IEEE Conference on Decision and Control*, 1985, pp. 49–51.
[8] T. L. Vu and K. Turitsyn, "Lyapunov Functions Family Approach to Transient Stability Assessment," *IEEE Trans. Power Syst.*, vol. 31, no. 2, pp. 1269–1277, Mar. 2016.
[9] H. K. Khalil, *Nonlinear Systems*, 3rd Edition. Pearson, 2002.
[10] S. P. Boyd and L. Vandenberghe, *Convex optimization*. Cambridge, UK ; New York: Cambridge University Press, 2004.
[11] C. Wang, V. A. Centeno, K. D. Jones, and D. Yang, "Transmission Lines Positive Sequence Parameters Estimation and Instrument Transformers Calibration Based on PMU Measurement Error Model," *IEEE Access*, vol. 7, pp. 145104–145117, 2019.